\documentclass[conference]{IEEEtran}
\IEEEoverridecommandlockouts
\usepackage{cite}
\usepackage{amsmath,amssymb,amsfonts}
\usepackage{algorithmic}
\usepackage{graphicx}
\usepackage{textcomp}
\usepackage{xcolor}
\usepackage{tikz}
\usepackage{float}
\usetikzlibrary{quantikz2}
\newcommand{\qg}[2]{\gate{#1\left(#2\right)}}  % quantum gate with argument
\newcommand{\rot}[2]{\qg{R_{#1}}{#2}}  % rotation gate with axis and argument
\newcommand{\raig}[1]{\rot{y}{\xi_{#1}}}  % rotation about axis a with indexed theta argument
\tikzset{
  mixed/.style={
    fill=none,
    append after command={
      \pgfextra{
        \begin{scope}
          \fill[black] (1.06pt, 1.06pt) arc (45:-135:1.5pt);
        \end{scope}
      }
    }
  }
}

\newtheorem{theorem}{Theorem}
\newtheorem{lemma}{Lemma}

\newcommand{\Endproof}{\hfill$\Box$\\}
\newcommand{\Beginproof}{{\em Proof}\quad\quad}
\def\BibTeX{{\rm B\kern-.05em{\sc i\kern-.025em b}\kern-.08em
    T\kern-.1667em\lower.7ex\hbox{E}\kern-.125emX}}
\begin{document}

\title{Quantum Random Forest for the Regression Problem
\thanks{The study was performed under the development programme of the Volga Region Mathematical Center (agreement no.  075-02-2026-1328).}
}
\author{\IEEEauthorblockN{Kamil Khadiev}
\IEEEauthorblockA{\textit{Institute of Computational Mathematics and IT} \\
\textit{Kazan Federal University}\\
Kazan, Russia \\
\textit{Zavoisky Physical-Technical Institute} \\
\textit{FRC Kazan Scientific Center of RAS}\\
Kazan, Russia \\
kamilhadi@gmail.com, http://orcid.org/0000-0002-5151-9908}
\and
\IEEEauthorblockN{Liliya Safina}
\IEEEauthorblockA{\textit{Institute of Computational Mathematics and IT} \\
\textit{Kazan Federal University}\\
Kazan, Russia \\
LiliyISafina@kpfu.ru, http://orcid.org/0000-0001-7182-3731}
}

\maketitle

\begin{abstract}
The Random Forest model is one of the popular models of Machine learning. We present a quantum algorithm for testing (forecasting) process of the Random Forest machine learning model for the Regression problem. The presented algorithm is more efficient (in terms of query complexity or running time) than the classical counterpart. 
\end{abstract}


\section{Introduction}
\emph{Quantum computing} \cite{nc2010,a2017,aazksw2019part1} is one of the hot topics in computer science of the last decades.
There are many problems where quantum algorithms outperform the best known classical ones \cite{quantumzoo},
and one of the most important performance metrics in this regard is \emph{query complexity}.
We refer to \cite{a2017,k2022lecturenotes} for a nice survey on the quantum query complexity.
%,and to \cite{aaksv2022,asa2024,aj2021,kkmsy2022,ki2019,kbcw2024,ke2022,kiv2022,kk2021,kb2022,kr2021b,kr2021a,kszm2022,ks2025,l2020,l2020conf,m2017} for the more recent progress on string processing problems.
One of the  hot topics in quantum computing is quantum machine learning \cite{QML1,aazksw2019part2,ssp2015qml,abbb2021qml,s2022qml,csk2024qml}. In this paper, we focus on the Random Forest model \cite{h1995rf,h1998rf,b2001rf,ag1997rf,hks1993rf} of machine learning. The model has a lot of different applications \cite{q2012,ks2019classical,bs2016,bjkk2012}. This is a supervised learning model that is an ensemble of decision tree models \cite{mr2014,cart,c50} using the bagging method \cite{b1996}.

There are two processes for a supervised machine learning model that are training (constructing a model by existing data with answers) and testing (computing an answer for a new input object). The training process for the Random Forest model is based on a decision tree training process. The quantum version of this process for decision trees can be found in \cite{m2023,kms2021,kms2019}. Quantum versions for the representation and usage of the decision tree model can be found in \cite{hbn2022,kylmp2025}.
The quantum versions of the testing (prediction or forecasting) process are discussed in \cite{skzk2023,ks2021,sj2025,shh2024,ksz2025} for Random Forest and ensample methods in general \cite{ks2022,s2023,ycx2024}. 

In this paper, we present a quantum algorithm for the testing (forecasting or prediction) process of the Random Forest model for the Regression problem. Let $n$ be the number of trees and $h$ be the maximum height of a tree. In that case, the complexity of the classical version of the testing algorithm is $O(n\cdot h)$ that is testing all the trees one by one. The suggested quantum version has $O(t\cdot h\cdot (y_{max}-y_{min}))$  query complexity, where $t$ is the required accuracy of the answer, $y_{max}$ is the maximum possible value for the answer, and $y_{min}$ is the minimal possible value of the answer. If it is enough to forecast a portion of the range between maximum and minimum values, then the query complexity is $O(t\cdot h)$.
Our method is based on the Quantum Amplitude Estimation algorithm \cite{bhmt2002} and the specific representation of the decision trees.

The structure of this paper is the following.
Section \ref{sec:prelims} describes the required notations and preliminaries. The quantum algorithm of testing (forecasting or prediction) for the  Random Forest model is presented in Section \ref{sec:rf}. The final Section \ref{sec:conclude} concludes the paper and contains some open questions. 

\section{Preliminaries}\label{sec:prelims}
\subsection{Basics of Machine Learning. Supervised Learning}
The supervised learning problem in Machine learning  \cite{ml1, ml2} allows us to forecast a result using information about existing objects. 
Let us consider the regression problem in a formal way. 

For a positive integer $m$, let ${\cal X}=(X^1, \dots, X^m)$ be a sequence of objects. An object $X^i=(x^i_1, \dots x^i_d)$ is a sequence of attributes for $i\in\{1,\dots,m\}$, where $d$ is a number of attributes. An attribute $x^i_j$ is a real-valued or a discrete-valued variable. In other words, for each $i\in\{1,\dots,m\}$, $x^i_j\in DOM_j$, where $DOM_j=\mathbb{R}$ if $x^i_j$ is a real-valued attribute; and $DOM_j=\{1,\dots,W_j\}$ if $x^i_j$ is a discrete-valued one.
Let ${\cal Y}=(y^1, \dots, y^m)$ be the sequence of results, where $y^i\in \mathbb{R}$.
The pair of $({\cal X}, {\cal Y})$ is called the training data.
Some of the known error functions are listed below. Let $\hat{y}^i$ be a forecast value for $X^i$:
%\begin{itemize}
 %   \item 
 $MAE=\frac{1}{m}\sum_{i=1}^m|y^i-\hat{y}^i|$,
  %  \item 
 $MSE=\frac{1}{m}\sum_{i=1}^m(y^i-\hat{y}^i)^2$,
   % \item 
 $RMSE=\sqrt{\frac{1}{m}\sum_{i=1}^m(y^i-\hat{y}^i)^2}$,
    %\item 
 $MAPE=\frac{1}{m}\sum_{i=1}^m\frac{|y^i-\hat{y}^i|}{|y^i|}$,    
    %\item 
 $wMAPE=\frac{\sum_{i=1}^m|y^i-\hat{y}^i|}{\sum\limits_{i=1}^m|y^i|}$,
    %\item 
 $sMAPE=\frac{1}{m}\sum_{i=1}^m\frac{|y^i-\hat{y}^i|}{|y^i|+|\hat{y}^i|}$.
%\end{itemize}

Using training data, we build a machine learning model that minimizes an error function. The building process is called training.
Then, we get a new input object  $X=(x_1,\dots,x_d)$. The goal is to compute a result $\hat{y}$ for $X$ that is called a forecast value. Here, we hope that the model minimizes the same error function for the forecast value $\hat{y}$ and the true value $y$. The process of computing $\hat{y}$ is called testing, forecasting or prediction.

In this paper, we assume that the model is already trained (built), and we are interested in the testing process only.
\subsection{Random Forest Model}
The training process for the Random Forest model can be found in \cite{h1995rf,h1998rf,b2001rf,ag1997rf,hks1993rf}. The main part of the process is to build decision trees, whose quantum algorithm can be found here \cite{m2023,kms2021,kms2019}. It is not the main topic of interest of the paper. Let us describe a ready trained model and the testing (forecasting) process.
Let us start with a decision tree model description that is a main part of the Random Forest model. 

A decision tree is a rooted tree. Each inner node tests some condition on input variables. Suppose $B$ is some test with outcomes ${b_1, b_2,\dots, b_g}$ that is tested in a node. Then, there are $g$ outgoing edges labeled by each result of the node's condition. A leaf is labeled with a real value.

The testing (forecasting) process is as follows. We start from the tree node. When we are in an inner node, we test the condition of the node and go by the edge labeled with the result of the condition. The testing process ends in a leaf. The leaf label is the result of the  process. 

Often, researchers consider binary trees, i.e. $g=2$. An example of a node condition for real-valued and discrete-valued attributes is  $``x_j>\theta_j``$, where $x_j$ is a value of the attribute $x_j$ and $\theta_j$ is a constant threshold. In that case, the possible outcomes are ``Yes'' or ``No''. The possible condition for discrete-valued attributes can be  $``x_j=\theta_j``$ with the same set of outcomes. 

As a Random Forest model, we consider a set of decision trees $\{T_0,\dots,T_{n-1}\}$, where $n$ is the number of trees. The parameter $h$ is the maximum height of the trees. Let us consider the testing (forecasting) process for an input object $X$. Let $\hat{y}_0,\dots, \hat{y}_{n-1}$ be the results that return trees for $X$. Then, the result of the testing (forecasting) process for the Random Forest model for $X$ is $\frac{1}{n}\sum_{i=0}^{n-1} \hat{y}_i$.
\subsection{Basics of Quantum Computing}
Quantum memory consists of quantum bits. A state of a qubit is a column-vector from ${\cal H}^2$ Hilbert space. It can be represented by $a_0|0\rangle+a_1|1\rangle$, where $a_0,a_1$ are complex numbers such that $|a_0|^2+|a_1|^2=1$, and $|0\rangle$ and $|1\rangle$ are unit vectors. Here we use the Dirac notation. A state of $N$ qubits is represented by a column-vector from ${\cal H}^{2^N}$ Hilbert space. It can be represented by $\sum\limits_{i=0}^{2^N-1}a_i|i\rangle$, where $a_i$ is a complex number such that $\sum\limits_{i=0}^{2^N-1}|a_i|^2=1$, and $|0\rangle,\dots |2^N-1\rangle$ are unit vectors. We have two types of transformation. The first one is a unitary transformation that is a multiplication of a state column vector to a unitary matrix $U$. The second one is called measurement. In that case, the quantum system collapses to one of the unit vectors $|j\rangle$ with probability $|a_j|^2$. 
Quantum unitary transformations are also called operators and gates.

Quantum circuits consist of qubits and a sequence of gates applied to these qubits. Graphically, on a circuit, qubits are presented as parallel lines. 
The reader can find more information about quantum circuits in \cite{nc2010,aazksw2019part1,k2022lecturenotes}.

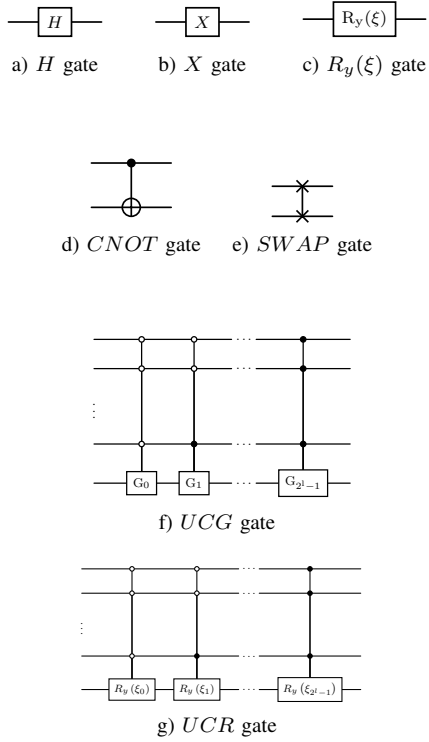
\begin{figure}[H]
    \centering
    \caption{Quantum gate circuits}
    \label{fig:qgates}
    \begin{table}[H]
    	\centering
    	\begin{center}
    		\begin{tabular}{ccc}
    			\begin{tikzpicture}
            	\node[scale=0.80] {
                	\begin{quantikz}
                    &\gate{H} & 
                	\end{quantikz} 
                    };
        		\end{tikzpicture}
    			 &
            \begin{tikzpicture}
            	\node[scale=.80] {
                	\begin{quantikz}
                    &\gate{X} & 
                	\end{quantikz} 
                    };
        		\end{tikzpicture}
    		
             & \begin{tikzpicture}
            	\node[scale=0.80] {
                	\begin{quantikz}
                    &\gate{\mathrm{R_{y}(\xi)}} & 
                	\end{quantikz} 
                    };
        		\end{tikzpicture}
    		\\
            a) $H$ gate & b) $X$ gate & c) $R_y(\xi)$ gate
    		\end{tabular}
    	\end{center}
    \end{table}
    \begin{table}[H]
    	\centering
    	\begin{center}
    		\begin{tabular}{cc}
    			\begin{tikzpicture}
            	\node[scale=0.80] {
                	\begin{quantikz}
                    &\ctrl{1} &\\
    		& \targ{} &
                	\end{quantikz} 
                    };
        		\end{tikzpicture}
                & 
                \begin{tikzpicture}
            	\node[scale=0.80] {
                	\begin{quantikz}
                    &\swap{1} &\\
    		& \targX{} &
                	\end{quantikz} 
                    };
        		\end{tikzpicture} 
    		\\
            d) $CNOT$ gate & e) $SWAP$ gate
    		\end{tabular}
    	\end{center}
    \end{table}
    \begin{table}[H]
    	\centering
    	\begin{center}
    		\begin{tabular}{c}
    		\begin{tikzpicture}
    	\node[scale=0.6] {
    		\begin{quantikz}
    		& \octrl{4} & \octrl{4} & \ \ldots\ & \ctrl{4}   &\\
    		& \octrl{3} & \octrl{3} & \ \ldots\ & \ctrl{3}   &\\
    			\wireoverride{1} \vdotswithin{\ldots}\\
    		  & \octrl{1} & \ctrl{1}  & \ \ldots\ & \ctrl{1}   &\\
    		& \gate{\mathrm{G_{0}}}  & \gate{\mathrm{G_{1}}}  & \ \ldots\ & \gate{\mathrm{G_{2^l-1}}} &
    		\end{quantikz}
    	       };
            \end{tikzpicture}
            \\ f) $UCG$ gate
    		
            \\
    
            \\
            
            \begin{tikzpicture}
    	\node[scale=0.5] {
    		\begin{quantikz}
    		& \octrl{4} & \octrl{4} & \ \ldots\ & \ctrl{4}   &\\
    		& \octrl{3} & \octrl{3} & \ \ldots\ & \ctrl{3}   &\\
    			\wireoverride{1} \vdotswithin{\ldots}\\
    		  & \octrl{1} & \ctrl{1}  & \ \ldots\ & \ctrl{1}   &\\
    		& \raig{0}  & \raig{1}  & \ \ldots\ & \raig{2^l-1} &
    		\end{quantikz}
    	       };
            \end{tikzpicture}  
            \\ g) $UCR$ gate
    		\end{tabular}
    	\end{center}
    \end{table}
\end{figure}

In the implementation (Section \ref{sec:impl}), we use several quantum gates, which are listed below:
%\begin{itemize}
   % \item 
    the Hadamard operator
 $H=\frac{1}{\sqrt{2}}\begin{pmatrix}
1 & 1 \\
1 & -1 
\end{pmatrix}$, 
%\item 
the inversion operator $X=\begin{pmatrix}
0 & 1 \\
1 & 0 
\end{pmatrix}$,
%\item 
the rotation operator
 $$R_y(\xi)=\begin{pmatrix}
cos(\xi/2) & -sin(\xi/2) \\
sin(\xi/2) & cos(\xi/2) 
\end{pmatrix},$$
%\item 
the controlled inversion or controlled not operator
$$ CNOT=\begin{pmatrix}
1  & 0 & 0 & 0\\
0  & 1  & 0 & 0\\
0  & 0  & 0 & 1\\
0  & 0  & 1 & 0
\end{pmatrix},$$
%\item 
the SWAP operator
$$SWAP=\begin{pmatrix}
1 & 0 & 0 & 0 \\
0 & 0 & 1 & 0 \\
0 & 1 & 0 & 0 \\
0 & 0 & 0 & 1
\end{pmatrix},$$
%\item 
the uniformly controlled gate (UCG) for $l-1$ controlled qubits
$$UCG = \left(
	\begin{array}{cccc}
		G_0 & 0 & \ldots & 0 \\
		0 & G_1 & \ldots & 0  \\
		\vdots & \ddots &\ddots & \vdots \\		
		0 & 0 & \ldots & G_{2^{l-1}} \\
	\end{array}
	\right),$$
where $G_i$ is some gate;
%\item 
the uniformly controlled rotation (UCR) for $l-1$ controlled qubits
$$UCR = \left(
	\begin{array}{cccc}
		R_0 & 0 & \ldots & 0 \\
		0 & R_1 & \ldots & 0  \\
		\vdots & \ddots &\ddots & \vdots \\		
		0 & 0 & \ldots & R_{2^{l-1}} \\
	\end{array}
	\right),$$
where $R_i$ is one-qubit rotation operator.
%\end{itemize}
The gate circuits are presented in Fig. \ref{fig:qgates}.

\subsection{Quantum Query Model}\label{sec:query-model}
One of the most popular computational models for quantum algorithms is the query model.
We use the standard form of the quantum query model. %is a generalization of the decision tree model of classical computation that is commonly used to lower bound the amount of time required by a computation. 
Let $f:D\rightarrow \{0,1\},D\subseteq \{0,1\}^M$ be an $M$ variable function. Our goal is to compute on an input $x\in D$. We are given an oracle access to the input $x$, i.e. it is implemented by a specific unitary transformation usually defined as $|i\rangle|z\rangle|w\rangle\rightarrow |i\rangle|z+x_i\pmod{2}\rangle|w\rangle$, where the $|i\rangle$ register indicates the index of the variable we are querying, $|z\rangle$ is the output register, and $|w\rangle$ is some auxiliary work-space. An algorithm in the query model consists of alternating applications of arbitrary unitary matrices which are independent of the input and the query unitary, and a measurement at the end. The smallest number of queries for an algorithm that outputs $f(x)$ with probability more than $ \frac{2}{3}$ on all $x$ is called the quantum query complexity of the function $f$ and is denoted by $Q(f)$.
We refer the reader to \cite{nc2010,a2017,aazksw2019part1,k2022lecturenotes} for more details on quantum computing. 
%
%In this paper, we are interested in the query complexity of the quantum algorithms. We use modifications of Grover's search algorithm \cite{g96,bbht98} as quantum subroutines. For these subroutines, time complexity can be obtained from query complexity by multiplication to a log factor \cite{ad2017,g2002}.   

\section{Quantum Algorithm for Testing Process of Random Forest}\label{sec:rf}
Before presenting the algorithm, let us list two facts from literature in Section \ref{sec:ae} that are used in our algorithm. Then, in Section \ref{sec:algo1} we present the algorithm for one decision tree and the algorithm for the Random Forest model in Section \ref{sec:algo2}.
\subsection{Tools}\label{sec:ae}
Let us recall two facts that will be useful for our algorithm. Let us consider an algorithm ${\cal U}$ that takes no input directly, but may have access to input during its execution using oracle queries (see the query model in Section \ref{sec:query-model}). Additionally, we assume that ${\cal U}$ is a quantum algorithm of the following form:
\begin{itemize}
    \item apply some unitary operator to the initial state $|0^N\rangle$ of $N$ qubits; 
    \item measure $K \leq N$ qubits of the resulting state
in the computational basis,
    \item obtaining outcome $r \in \{0, 1\}^K$;
    \item output $f(r)$ for some easily computable
function $f : \{0, 1\}^K \to \mathbb{R}$.
\end{itemize}
 We finally assume that we have access to the inverse of the unitary part  of the algorithm ${\cal U}^{-1}$.
 
 The first fact is simple and well known. Sometimes, it is called the powering lemma or success probability boosting technique.

\begin{lemma}[Powering lemma \cite{jvv1986} or success probability boosting technique]\label{lm:pl} Let ${\cal U}$ be a (classical or quantum) algorithm which aims to
estimate some quantity $\beta$, and whose output $\tilde{\beta}$ satisfies $|\beta-\tilde{\beta}|\leq \varepsilon$ except with probability $\gamma$ , for some
fixed $\gamma<0.5$. Then, for any $\delta>0$, it suffices to repeat the  ${\cal U}$ algorithm $O(\log \frac{1}{\delta})$ times and take the median to obtain an
estimate which is accurate to within $\varepsilon$ with probability at least $1-\delta$.
\end{lemma}
Examples of application of the lemma for quantum algorithms can be found in \cite{ks2019,ks2023,kks2019,kksk2020,m2015}.
%ссылки на работы об использование леммы)

The second fact is the amplitude estimation algorithm.

\begin{lemma}[Amplitude estimation \cite{bhmt2002}]\label{lm:ae}
 There is a quantum algorithm called amplitude estimation which takes as input one copy of a quantum state $|\Psi\rangle$, a unitary transformation $D=2|\Psi\rangle\langle\Psi|-I$, a unitary transformation $V=I-2P$ for some projector $P$, and an integer $t$. The algorithm outputs $\tilde{\beta}$, an estimate of $\beta=\langle\Psi|P|\Psi\rangle$, such that 
 \[|\beta-\tilde{\beta}|\leq 2\pi \frac{\sqrt{\beta(1-\beta)}}{t}-\frac{\pi^2}{t^2}\]
   with probability at least $\frac{8}{\pi^2}$, using $D$ and $V$ transformations $t$ times each. 
\end{lemma}

The success probability of $\frac{8}{\pi^2}$ can be improved to $1-\delta$ for any $\delta>0$ using the powering lemma at the cost of a multiplicative factor $O(\log\frac{1}{\delta})$.

\subsection{Algorithm for One Decision Tree}\label{sec:algo1} 
Let us start from the description of processing a single tree  $T_i\in\{T_0,\dots,T_{n-1}\}$.

Let $y_{min}$ and $y_{max}$ be the minimal and maximum possible labels of leaves among all the trees of the model. Formally, \[y_{min}=\min\{y[0,1],\dots,y[0,k_0],\]\[y[1,1],\dots,y[1,k_1],\dots,y[n-1,1],\dots,y[n-1,k_{n-1}]\},\]
\[y_{max}=\max\{y[0,1],\dots,y[0,k_0],\]\[y[1,1],\dots,y[1,k_1],\dots,y[n-1,1],\dots,y[n-1,k_{n-1}]\}.\]
Here, $k_i$ is the number of leaves of a tree $T_i$, and $y[i,1],\dots,y[i,k_i]$ are labels of the leaves.

Let us enumerate all nodes of the tree $T_i$ from $1$ to $2^h-1$ level by level from the root to the leaves. On each level from left to right.

Let us consider three quantum registers:
\begin{itemize}
    \item $|\lambda\rangle$ of $\log_2 n$ qubits is an index of the tree.
    \item $|\psi\rangle$ of $h+1$ qubits is an index of a node.
    \item $|\phi\rangle$ of $1$ qubit is for the result of computation.
\end{itemize}

For processing the tree, we perform the following steps.
\begin{itemize}
    \item[] \textbf{Step 1.} We choose the leaf. Here we start from the state $|\lambda\rangle|\psi\rangle|\phi\rangle=|i\rangle|1\rangle|0\rangle$, emulate the deterministic process of walking by the tree $T_i$ from the root to a leaf according to the conditions and reach a leaf-node with index $j_i$.
    
    \item[] \textbf{Step 2.} In this step, we are in the note $j_i$ of the tree $i$. Let the leaf be labeled with the value $y_i=y[i,j'_i]$, where $j'_i$ is the index of the corresponding value of the tree $T_i$. Then, we rotate the qubit $|\phi\rangle$ to the angle
    \[\alpha_i=arcsin\Bigg(\sqrt{\frac{y_i-y_{min}}{y_{max}-y_{min}}} \Bigg).\]

    Because $0\leq \frac{y_i-y_{min}}{y_{max}-y_{min}}\leq 1$, we have $\left(\sin \alpha_i\right)^2=\frac{y_i-y_{min}}{y_{max}-y_{min}}$.

    For this goal, we apply the operator $R_y(2\alpha_i)$ \[|i\rangle|j_i\rangle|0\rangle\to|i\rangle|j_i\rangle\left(\sin\alpha_i|1\rangle + \cos\alpha_i|0\rangle\right).\]
\end{itemize}

Let us denote the described procedure as $U_{T_i}$.
\subsection{Algorithm for the Random Forest Model}\label{sec:algo2} 
Let us describe the algorithm for the whole forest. The algorithm has two phases:
\begin{itemize}
    \item[] \textbf{Phase 1.} We start with the register $|\lambda\rangle|\psi\rangle|\phi\rangle=|0\rangle|0\rangle|0\rangle$. Firstly, we apply the Hadamard operator $H$ to each qubit of $|\lambda\rangle$ and set the register $|\psi\rangle$ equal to $|1\rangle$:
    %\[
    $|0\rangle|0\rangle|0\rangle\to\frac{1}{\sqrt{n}}\sum_{i=0}^{n-1}|i\rangle|1\rangle|0\rangle.
    $
    %\]
    \item[] \textbf{Phase 2.} We apply the transformation $U$ which is an application $U_{T_i}$ if  $|\lambda\rangle=|i\rangle$.

\vspace{-0.5cm}
\[
\frac{1}{\sqrt{n}}\sum_{i=0}^{n-1}|i\rangle|1\rangle|0\rangle\to\frac{1}{\sqrt{n}}\sum_{i=0}^{n-1}|i\rangle|j_i\rangle\left(\sin\alpha_i|1\rangle + \cos\alpha_i|0\rangle\right)
\]
\end{itemize}
Let us denote the algorithm as ${\cal U}$. So, if we apply it to $|\lambda\rangle|\psi\rangle|\phi\rangle=|0\rangle|0\rangle|0\rangle$, then obtain a state $|\Psi\rangle={\cal U}|0^{\log_2n+h+2}\rangle$. Let us define the projector $P=I\times |1\rangle\langle 1|$. The transformation $D$ can be implemented as
\[D=2|\Psi\rangle\langle \Psi|-I=2{\cal U}|0^{\log_2n+h+2}\rangle\langle 0^{\log_2n+h+2}|{\cal U}^{-1}-I.\]
The transformation $V=I-2P$.
Using transformations $D$ and $V$, we can apply the amplitude estimation algorithm (Lemma \ref{lm:ae}) and the powering lemma (Lemma \ref{lm:pl}) with respect to the required accuracy $t$ and the error probability $\delta$.

As a result, the algorithm returns the value $\tilde{\beta}$ that estimates $\beta$ with accuracy $t$ and error probability $\delta$. The value $\beta$ is
\[\beta = \frac{1}{n}\sum_{i=0}^{n-1}\frac{y_i-y_{min}}{y_{max}-y_{min}}=\]\[ \frac{1}{(y_{max}-y_{min})}\cdot\frac{1}{n}\sum_{i=0}^{n-1}y_i - \frac{1}{(y_{max}-y_{min})}\cdot\frac{1}{n}\sum_{i=0}^{n-1}y_{min}=\]\[
\frac{1}{(y_{max}-y_{min})}\cdot R + C \]
Here, $R=\frac{1}{n}\sum\limits_{i=0}^{n-1}y_i$ is the target result and \[C=\frac{1}{(y_{max}-y_{min})}\cdot\frac{1}{n}\sum\limits_{i=0}^{n-1}y_{min}=\]\[\frac{1}{(y_{max}-y_{min})}\cdot\frac{n\cdot y_{min}}{n}=\frac{y_{min}}{(y_{max}-y_{min})}.\] $C$ is the constant that does not depend on the input object.

We can compute the target value as $R=(\beta-C)(y_{max}-y_{min}).$
This means that we should increase the accuracy for $\beta$, and it should be $t(y_{max}-y_{min})$. In that case, we reach the accuracy $t$ for the target value $R$. Because of the amplitude estimation algorithm's complexity, it is also affect the complexity of the algorithm.
At the same time, if it is enough for us to compute a proportion of the range between the maximum and the minimum, then we compute $\beta$, and we can use only accuracy $t$.
Depending on the procedure for checking quality of forecasting, we can choose computing $\beta$ or $R$. If we use procedures similar to $MAE$, $MSE$, $RMSE$ (difference of the values), then we should compute $R$. If we use procedures similar to  $MAPE$, $wMAPE$, $sMAPE$ (a portion of values), then we can compute $\beta$.% We discuss it in the next lemma
%\begin{lemma}\label{lm:mape}   MAPE, PMAD or SS \end{lemma}
%\Beginproof \Endproof

Let us present the analysis of the presented algorithm in the following theorem.
\begin{theorem}
The presented quantum algorithm returns the result of the testing process of the Random Forest model for an object $X$ with accuracy $t$ and error probability $\delta$. The query complexity of the algorithm is $O(t\cdot h\log\frac{1}{\delta})$ for forecasting $\beta$ value, and is $O(t(y_{max}-y_{min})\cdot h\log\frac{1}{\delta})$ for forecasting $R$ value.
\end{theorem}
\Beginproof
As we discussed before, the application of the algorithm ${\cal U}$ to the $|\lambda\rangle|\psi\rangle|\phi\rangle=|0\rangle|0\rangle|0\rangle$ state gives us the state $|\Psi\rangle=\frac{1}{\sqrt{n}}\sum\limits_{i=0}^{n-1}|i\rangle|j_i\rangle\left(\sin\alpha_i|1\rangle + \cos\alpha_i|0\rangle\right)$.
We have
\[\langle \Psi|P|\Psi\rangle=\frac{1}{n}\sum_{i=0}^{n-1}(\sin(\alpha_i))^2=\]\[\frac{1}{n}\sum_{i=0}^{n-1}\left(\sin\left(arcsin\left(\sqrt{\frac{y_i-y_{min}}{y_{max}-y_{min}}}\right)\right)\right)^2=\]\[\frac{1}{n}\sum_{i=0}^{n-1}\frac{y_i-y_{min}}{y_{max}-y_{min}}=\beta.\]

Next, we apply the amplitude estimation algorithm (Lemma \ref{lm:ae}), and obtain an estimate $\tilde{\beta}$ for $\beta=\frac{R}{(y_{max}-y_{min})} + C$ such that
$|\tilde{\beta}-\beta|\leq 2\pi\frac{\sqrt{\beta(1-\beta))}}{t}+\frac{\pi^2}{t^2}\leq \frac{2\pi}{t}+\frac{\pi^2}{t^2}$.
The last inequality is correct because $0\leq\frac{y_i-y_{min}}{y_{max}-y_{min}}\leq 1$, and $0\leq\frac{1}{n}\sum\limits_{i=0}^{n-1}\frac{y_i-y_{min}}{y_{max}-y_{min}}\leq 1$. Recall that $\beta=\frac{R}{(y_{max}-y_{min})} + C=\frac{1}{n}\sum\limits_{i=0}^{n-1}\frac{y_i-y_{min}}{y_{max}-y_{min}}$. So, we obtain $0\leq \beta\leq 1$, and $0\leq 1-\beta\leq 1$. Hence, $0\leq \beta(1-\beta)\leq 1$.
We estimate $\beta$ by $\tilde{\beta}$ with probability at least $8/\pi^2$. The power lemma (Lemma \ref{lm:pl}) implies that the median of $O(\log \frac{1}{\delta})$ repetitions will lie within this accuracy bound with probability $1-\delta$.

Let us discuss the query complexity of the algorithm. The amplitude estimation algorithm invokes $D$ and $V$ transformations $t$ times. The transfomration $V$ does not depend on input variables, but $D$ does. The transformation $D$ invokes ${\cal U}$ and ${\cal U}^{-1}$ once. Each of them depends on $U$ that do $O(h)$ queries on each node (we test an attribute as an  input variable to compute a condition) from roots to leaves.  So, the complexity of $D$ is $O(h)$, and the complexity of the amplitude estimation algorithm is $O(t\cdot h)$. We invoke it $O(\log \frac{1}{\delta})$ times because of the powering lemma. The total complexity is $O(t\cdot h \log \frac{1}{\delta})$.

If we estimate $R=(\beta-C)(y_{max}-y_{min})$, then we need to increase accuracy. It becomes $t(y_{max}-y_{min})$. So, the complexity of the amplitude estimation algorithm is $O(t(y_{max}-y_{min})\cdot h)$, and the total complexity is $O(t(y_{max}-y_{min})\cdot h \log \frac{1}{\delta})$.
\Endproof

\section{On Circuit Implementation}\label{sec:impl}
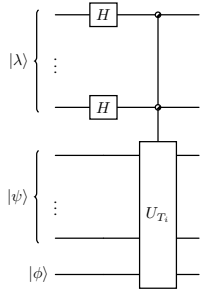
\begin{figure}[H]
    \centering
    \caption{Quantum random forest forecasting circuit}
    \label{fig:qrf}
    \begin{tikzpicture}
        \node[scale=0.6] {
            \begin{quantikz}
                \lstick[3]{\ket{\lambda}}    
                & \gate[1]{H} & \ctrl[style={mixed}]{2}& \\
                \wireoverride{1} \vdotswithin{\ldots}\\
                 & \gate[1]{H}& \ctrl[style={mixed}]{1} &\\
                \lstick[3]{\ket{\psi}} 
                & & \gate[4]{U_{T_i}} &\\
                \wireoverride{1}\vdotswithin{\ldots}\\
                &  &  & \\
                \lstick{\ket{\phi}}    
                & & & 
            \end{quantikz}
        };
    \end{tikzpicture}
\end{figure}

The quantum circuit for the random forest forecasting algorithm described in Section \ref{sec:algo2} (the operator  ${\cal U}$) is presented in Fig. \ref{fig:qrf}.

The gate $U_{T_i}$ is $UCG$ controlled by the decision tree indices in the register $|\lambda\rangle$. The transformation is described in section \ref{sec:algo1}. Step 1 consists of a sequence of multiplication by 2 and addition of 1 operations. The register $|\psi\rangle$ updates as follows: $|j\rangle \to |j\times 2\rangle$ if the condition at the corresponding node is false; otherwise, $|j\rangle = |j\times 2 + 1\rangle$. This process is repeated $h$ times. 

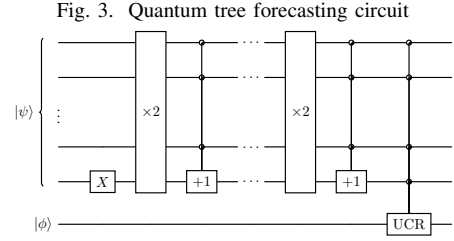
\begin{figure}[H]
    \centering
    \caption{Quantum tree forecasting circuit}
    \label{fig:qtree}
    \begin{tikzpicture}
        \node[scale=0.55] {
            \begin{quantikz}
                \lstick[5]{\ket{\psi}}    
                & & \gate[5]{\mathrm{\times 2}} & \ctrl[style={mixed}]{4} & \ \ldots\ & \gate[5]{\mathrm{\times 2}} & \ctrl[style={mixed}]{4} & \ctrl[style={mixed}]{5} & \\
                & &  & \ctrl[style={mixed}]{3} & \ \ldots\ & & \ctrl[style={mixed}]{3} & \ctrl[style={mixed}]{4} &\\
                \wireoverride{1} \vdotswithin{\ldots}\\
                 & & & \ctrl[style={mixed}]{1}  & \ \ldots\ & & \ctrl[style={mixed}]{1}  & \ctrl[style={mixed}]{2} & \\
                & \gate{X}& & \gate[1]{\mathrm{+1}} & \ \ldots\ & & \gate[1]{\mathrm{+1}}&  \ctrl[style={mixed}]{1} & \\
                \lstick{\ket{\phi}}    
                 & & & & & & & \gate{\mathrm{UCR}}  &
            \end{quantikz}
        };
    \end{tikzpicture}
\end{figure}

Step 2 is implemented using uniformly controlled rotation operator.
The quantum circuit for the operator $U_{T_i}$ is presented in Fig. \ref{fig:qtree}.

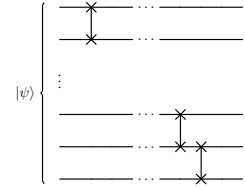
\begin{figure}[H]
    \centering
    \caption{The operator of multiplication by 2 circuit}
    \label{fig:qmultby2}
    \begin{tikzpicture}
        \node[scale=0.55] {
            \begin{quantikz}
                \lstick[6]{\ket{\psi}}    
                &\swap{1} &  \ghost{} & \ \ldots \ &  \ghost{} &  \ghost{} & & 
                \\
               & \targX{} & & \ \ldots \ &  \ghost{} &  \ghost{} & &
                \\
               %&   & \targX{}  & \ \ldots\ &  \ghost{} & \ghost{} & & \\
                \wireoverride{1} \vdotswithin{\ldots}\\
                & \ghost{} &  \ghost{} & \ \ldots\ & \swap{1} & \ghost{} & &\\
                & \ghost{} & \ghost{} & \ \ldots\ & \targX{} & \swap{1} & &\\
                & \ghost{} & \ghost{} & \ \ldots\ & \ghost{} & \targX{} & &
            \end{quantikz}
        };
    \end{tikzpicture}
\end{figure}

The operator of multiplication by 2 is a sequence of SWAP gates due to \cite{skzk2023}. Its circuit is in Fig. \ref{fig:qmultby2}.

The ''+1'' operation adds one by applying an inversion operator to the least significant qubit of the register $|\psi\rangle$ when the parent node's condition is met. The operator is controlled by the qubits of the register $|\psi\rangle$. It is implemented using $UCG$.

If we are interested in representation of the circuit using basic gates ($H, R_y, X, CNOT$), then the SWAP gate can be represented using $3$ CNOT gates. Representation of UCG and UCR gates can be found in \cite{mottonen2006decompositions,zk2025,zk2025icmne,ksy2024}.

For testing the quantum circuit, we generate a random forest with randomized parameters consisting of two trees, each of height $2$. Small values are selected due to the limitations imposed by simulator performance. The accurance $t = 32$, equivalent to 5 qubits. We use two ancilla qubits and a register storing an object $X$. For simplicity, $X$ contains $3$ binary attributes. The total number of qubits equals $14$ --- substantial for standard computers simulated a quantum circuit. The circuit executes 10 iterations. Our experiments show that with high probability we reach the required accuracy, and the results differs from the classical algorithm only on about $3\%$ for the value $R$. For the normilized value $\beta$, examples of results are $0.1596$ for classical implementation and $0.14645$ for quantum implementation. According to Lemma 2, the difference between normalized predictions should not exceed $0.0623$, making this test highly successful.
However, probabilistic measurement processes can occasionally produce less accurate outcomes.
%For example, when $X = 011$, the classical prediction yields 16.4554 (normalized to 0.0138), but the quantum version outputs $21.728$ (normalized to $0.0381$) --- an acceptable though suboptimal result. The obtained result does not meet the desired level of precision, however, it is still acceptable.
\section{Conclusion}\label{sec:conclude}
In the paper, we suggest a quantum algorithm for testing (forecasting) process of the Random Forest model for the Regression problem. The query complexity of the algorithm do not depend on the number of trees, but depend on the accuracy of computation. If we forecast a portion of the range between the maximum and the minimum, then the complexity is $O(t\cdot h)$, where $t$ is an accuracy of computation and $h$ is the height of trees. If we forecast the value itself, then the complexity is $O(t(y_{max}-y_{min})\cdot h)$.
The open question is how to remove dependence on $(y_{max}-y_{min})$.
%\section*{Acknowledgment}
%We thank ....

\bibliographystyle{unsrt}
\bibliography{tcs}

\end{document}